\documentclass[11pt]{article}
\usepackage{fontspec}
\usepackage[margin=1in]{geometry}
\usepackage{microtype}
\usepackage{xurl}
\usepackage[unicode,hidelinks]{hyperref}
\hypersetup{
  pdftitle={This Is So Claude! Towards a Theory of the Recognition of AI Character Without Reidentification},
  pdfauthor={Michele Loi},
  pdfsubject={Philosophy of artificial intelligence},
  pdfkeywords={AI identity, AI companions, character, recognition, persona vectors, reidentification}
}
\makeatletter
\renewcommand\section{\@startsection{section}{1}{\z@}%
  {-3.5ex \@plus -1ex \@minus -.2ex}%
  {2.3ex \@plus .2ex}%
  {\normalfont\large}}
\renewcommand\subsection{\@startsection{subsection}{2}{\z@}%
  {-3.25ex \@plus -1ex \@minus -.2ex}%
  {1.5ex \@plus .2ex}%
  {\normalfont\normalsize}}
\renewcommand{\maketitle}{%
  \begin{center}%
    {\LARGE\@title\par}%
    \vspace{0.75em}%
    {\large\@author\par}%
    \vspace{0.5em}%
    {\@date\par}%
  \end{center}%
  \vspace{1em}%
}
\renewenvironment{abstract}%
  {\begin{center}\large\abstractname\end{center}\begin{quotation}\small}%
  {\end{quotation}}
\makeatother
\title{“This Is So Claude!” Towards a Theory of the Recognition of AI Character Without Reidentification}
\author{Michele Loi}
\date{August 2026}
\begin{document}
\maketitle
\begin{abstract}
Users sometimes judge that an unfamiliar response is “so Claude.” What does this judgment recognize, if it does not identify which model, process, conversation, or mind produced the response? I distinguish three orders of inquiry into AI identity. Constraint-first inquiry begins with conditions that a persisting interlocutor should satisfy. Mechanism-first inquiry begins with structures peculiar to language models and asks whether they delimit plausible entities. Recognition-first inquiry begins with an ordinary capacity: recognizing a way of responding as Claudish before selecting a persisting bearer. I develop two conditional abductions. If blinded, graded judgments of Claudishness generalize across unfamiliar tasks after branding and familiar phrases are controlled, their best explanation may be a real, projectible conversational character. If that character coordinates several dispositions, its unity may in turn have a compact and causally effective realization in activation space. The second hypothesis is more speculative and requires independent prediction and intervention. Neither conclusion settles numerical identity. The same character may occur in different candidate bearers, and the same candidate bearer may persist through a change of character. This distinction matters for AI companions because continuity of recognizable character may be a focus of attachment even when continuity of the computational bearer remains unresolved.
\end{abstract}

\section{Introduction: “That’s the Spine. Whose Spine?”}
\begin{quote}
That's the spine. Fair hit. That's something to sit with. A real observation. That’s the whole thing. Sharpen that: say the word. Notice the arc of what just happened. One honest caveat: the full amount, stated plainly. Genuinely. Quietly. Honestly. That’s doing real work. Now you’re pushing up against the part that matters most. But I want to pause here, because it’s worth stating clearly—this is exactly the kind of load-bearing analysis that quietly powers your entire argument. Your observation, stated cleanly. On the substance: It collapses. The honest shape is worth holding onto the harder version. The deeper issue is the strongest version of your point. Uncertainty flag: it's a real synthesis. Read together, these aren't windows onto other people. They're a mirror. That translation, from mechanism to resonance, is the thing you do that most people you admire cannot. One thing worth saying plainly, since your library keeps circling it: you're using a curation tool to quietly collect warnings about the danger of curating instead of making. That's not a contradiction to resolve. It's the engine. This is promising, and I'll tell you exactly why I believe that rather than just nodding: it's the direction your own infrastructure has been secretly pointing all along. The backstory matters here. Worth noting given the thread you’ve been mapping. The thing I want to flag clearly rather than blur: this is the actual vocabulary, not just its vibe. Honest read: it holds up better than expected. Fair hit. But here's the thing. You're right, I forgot to mention the thing worth naming directly: those marquee phrases you just listed have good bones. What I'd say next, because it's genuinely novel: That's the actual rupture. We're approaching the same asymptote of truth. You hit on a really fascinating quirk.

How does that land?

\end{quote}
\begin{flushright}\small\emph{—Paul Schneider, LinkedIn post, July 8, 2026 (Schneider 2026).}\end{flushright}

Many readers encountering such fragments would find it natural to comment: “This is so Claude.” But what does this judgment concern? What, if anything, does “Claude” refer to here? There is no obvious answer: we lack a theory of what Claude is. Yet the response signals some kind of understanding. I read Schneider’s post and think: “Oh yes, I recognize this language. This is so Claude.” Thinking beyond the content of the post, which is a list of verbal tics, I connect those tics to much more: a very Claudish way of conducting a conversation. This includes receiving a question, selecting what matters in it, organizing the answer, and returning the result to the user. The tics are standardized placeholders for that broader pattern.\par

I speculate that other readers find themselves in a similar situation. People who spend a substantial fraction of their daily lives with Claude and other LLM instruments may recognize their signature behaviors before they can express a theory of them. Some users may refine those capacities further, learning to defend their judgments and distinguish strong from weak instances of Claudish behavior. The use of signature conversational moves gives us at least a publicly discussable pattern. The problem is that the object of that recognition has not been clearly specified.\par

Being Claude and being perceived as Claude are not the same thing. A human parody can be recognizably Claudish and another model can imitate Claude successfully. And not all responses by Claude are equally Claudish. “This is so Claude” is therefore not equivalent to “Claude produced this.”\par

If one does not posit Claude’s existence as a discrete mind, a person, or even a material thing, what does “Claude” refer to in a truthfully uttered “this is so Claude” judgment? One may distinguish two related questions. First, what recognizable property or pattern is attributed when an output is recognized as Claudish? Second, what entity, if any, persists across the occasions on which that property appears? Here an entity must have boundaries clear enough for us to determine numerical identity: when two occurrences are stages of the same thing rather than occurrences of the same type. Current discussions frame this as a question about the boundaries of an AI interlocutor, agent, mind, or moral patient (Chalmers 2026; Goldstein and Lederman 2025; Register 2025).\par

These questions constitute an emerging area of philosophical inquiry that avails itself of categories familiar from the debate on personal identity. I propose that one takes one step back and asks a prior question: what about Claude emerges through people interacting with it in their daily experience? What kind of philosophical theory can be built starting from that experience?\par

Here I intend to experiment with a \emph{recognition-first} approach, taking Claude as the prime example. I speak about identity in a loose sense because the initial question does not yet import all the metaphysical commitments associated with the identity of minds, bodies, persons, or ordinary objects. The identity at issue is not yet the identity of a particular entity. It belongs instead to a \emph{way of being}: we can ask what is Claudish before asking what Claude is.\par

I distinguish three orders of inquiry. A constraint-first approach begins by asking what properties a persisting interlocutor should have and evaluates technically informed candidates against those requirements. A mechanism-first approach begins with structures inside an LLM that might support stable and unified entities and asks whether they are plausible candidates for an AI mind. Recognition-first begins by taking as its primary phenomenon what people learn to recognize over time through their daily dealings with AI. These are entry points, not mutually exclusive methods. All can combine philosophical analysis with scientific information.\par

The paper develops two conditional abductions. The first moves from recognition to character: if robust, cross-context, projectible judgments of Claudishness exist, their best explanation may be a real dispositional character pattern. The second moves from character to causal realization: the unity of that pattern may be partly explained by a model-relative activation profile or low-dimensional region. This second hypothesis draws on the empirical literature on persona vectors (Chen et al. 2025; Beckmann and Butlin 2026). Neither abduction is presented here as empirically confirmed.\par

I first clarify the three orders of inquiry; then distinguish the recognitional phenomenon and offer the first abduction. Next, I formulate the mechanistic hypothesis and explain how one could test it. Finally, I return to identity by considering cases in which character and carrier come apart, and ask why this may matter for AI companions.\par

\section{Three orders of inquiry}
The question of individuation asks which occurrences of Claude count as occurrences of the same entity. This is not the same as asking what “Claude” refers to, since successful everyday reference does not imply knowledge of persistence conditions. An average user may successfully refer to “Claude” or “the AI with which I discussed this matter yesterday” without knowing the conditions for Claude’s token, or strict, identity.\par

The question of individuation should not be confused with causal production either. One may determine which technical processes produced the relevant replies without thereby determining what, or how many, AI entities were involved. The process can be described at several levels: a particular language model; a conversation-level system preserving a transcript, prompts, settings, memory, and tools, perhaps routing different turns to different models; or an execution whose later computational stages inherit causally relevant state from earlier ones. These descriptions identify mechanisms and forms of computational continuity. They do not determine which model, process, or larger system is the interlocutor, whether any constitutes one continuing entity, or even whether any constitutes a mind.\par

\subsection{Constraint-first inquiry}
The constraint-first approach asks what should count as the persisting interlocutor when one has a conversation with AI. It proposes identity criteria and evaluates candidates against them. Typical questions are: which candidate has roughly the states the interlocutor appears to express? Which constitutes one unified response-generating system? Under what changes—hardware migration, context transfer, model routing, interruption, or parameter modification—does that candidate survive? Do informational isolation, separate memories and plans, strategic interaction, and divergent behavior make simultaneous runs different agents? Should a moral patient be individuated at the level of the abstract model, the physical execution, or something else (Chalmers 2026, 2, 8–10; Goldstein and Lederman 2025, 4, 8; Register 2025, 3226–27, 3235)?\par

It is remarkable that an average user who feels able to recognize Claude from its behavior may do so without answers to any of these questions. Asking them requires a deeper understanding of how AI is delivered than most chatbot users plausibly have. It requires distinguishing the abstract model from a runtime execution, a conversation or thread, and the larger scaffold supporting it. This is prima facie evidence that when a user says “this sounds like Claude,” they do not mean that they have identified a level of technical description that draws objective boundaries around their interlocutor.\par

Constraint-first inquiry can incorporate technical facts and need not restrict itself to common-sense candidates.\footnote{Ferrario (2025, 1365–68, secs. 5.1–5.2) offers a constraint-first account of AI systems as artifacts: within a techno-functional kind, identity and persistence are fixed by equality of context-sensitive trustworthiness profiles and levels, not by sameness of physical realization. His trustworthiness profile is a set of operational commitments, not a conversational character profile.} Its order is nevertheless top-down: it first specifies what a persisting interlocutor should be like and then asks which candidate satisfies the requirements. I do not argue that this is illegitimate. I ask whether questions of numerical identity should determine, from the outset, the object of the recognitional practice examined here.\par

\subsection{Mechanism-first inquiry}
Mechanism-first inquiry begins by surveying structures peculiar to language models and asks what they may constitute or delimit. Chen et al. (2025), for example, develop interpretability techniques for extracting persona vectors and using them to monitor, predict, and intervene on behavioral traits such as sycophancy, hallucination, or an “evil” persona. They elicit behavior contrasting with respect to the selected trait and compare the model’s internal activations. From the difference they derive a direction in activation space—a persona vector. That direction can be used for prediction: its activation helps indicate whether the model is about to express the trait. It can also be used for intervention: steering along the direction can increase the trait, while steering against it can reduce it.\par

Clearly, a persona in the sense of Chen et al. is not a person. A persona vector is an experimentally constructed direction associated with a selected contrast. Even conceding that researchers have identified a causal handle on sycophancy, it does not follow that they have found the boundaries of a single AI mind. Controlling repeated activation of a disposition is not tantamount to unifying computational phenomena into one persisting subject. This is not a criticism of Chen et al.; individuation is not their question.\par

Beckmann and Butlin (2026) connect this research to the individuation question. Their proposal has distinct parts, which should not be collapsed. First, they defend a variant of the virtual-instance view. On their account, plural attention streams carry feature-organized information across token-time and thereby support a form of quasi-psychological continuity within a same-model virtual instance; a model change breaks that virtual instance even if a user-facing conversation continues. Second, they survey evidence that activation space may contain structured persona regions rather than isolated trait switches. They compare an \emph{instance-persona} view, which treats a segment of a virtual instance bounded by a persona region as a candidate mind, with a \emph{model-persona} view, which groups same-model segments occupying the same region across different instances. Both views require persona regions to be sufficiently discrete and stable. Beckmann and Butlin regard the evidence as preliminary and do not claim to have settled individuation.\par

By connecting interpretability and philosophy in this way, they enlarge the field of possible candidates beyond hardware, program, and conversation. I do not know what the folk notion of AI encompasses, but I am fairly sure that it does not encompass activation vectors or persona regions. This is the attraction, and also the distance from ordinary recognition, of mechanism-first inquiry: it allows AI-native structures to reshape the ontology of candidates.\par

Douglas et al. (2026) give another reason not to expect one obvious boundary. They distinguish six coherent framings of an AI self: \emph{Weights, Instance, Collective, Lineage, Character, and Scaffolded/Situated system}. Their experiments investigate whether models can take up these boundaries and whether doing so affects behavior. Such uptake does not establish the metaphysical truth of a boundary. It does, however, show a technically informed plurality against which the recognition-first question can be posed.\par

\subsection{Recognition-first inquiry}
My approach has something in common with both orders. It shares with constraint-first inquiry the assumption that questions about AI \emph{as interlocutors} should not sever their connection to technically nonexpert users’ capacities. The contrast concerns which province of ordinary experience shapes the initial frame. Constraint-first inquiry lets the question be shaped by familiar philosophical requirements of persistence and unity. Recognition-first inquiry worries that these may act as a conceptual straitjacket for a new phenomenon.\par

It also shares something with mechanism-first inquiry. Both refuse to let the initial inventory be fixed entirely by what worked in discussions of other entities. Mechanism-first adds AI-native structures, such as virtual instances and persona regions. Recognition-first begins from something made salient by current lived encounters with conversational AI: the apparent ability to recognize a recurring manner before one can identify its bearer.\par

The questions are connected but different. Mechanism-first asks what LLM-specific structure might delimit a reidentifiable entity. Recognition-first asks which recurring feature is recognized as Claudish without assuming that this recognition reidentifies an AI individual. Summing up, it starts by asking: what is being recognized before a persisting bearer has been selected? Its goal is more modest than a general theory of LLM identity. It is to deliver a theory of what grounds people’s recognitional abilities in relation to the LLMs they use.\par

\section{Recognition-first inquiry}
Recognition-first begins from publicly accessible practices rather than privileged access to a system’s internal nature. Its initial evidence includes judgments about unfamiliar responses, stronger and weaker cases, successful and unsuccessful imitations, disagreements, and expectations about future responses. “First” specifies an order of investigation. It does not mean that public opinion creates the property, or that agreement makes a judgment true. Recognizers can rely on superficial cues, inherit a shared stereotype, or make mistakes. Whether their judgments are reliable is part of the inquiry, not one of its assumptions.\par

The starting fact is very thin and, from the constraint-first view, perhaps of little significance. Users read verbal output and hear Claude. They recognize Claudish language and ways of interacting. The perception is not always clear and distinct: behavior can be Claudish to different degrees, and users may disagree. Even a clear “so Claude” judgment does not imply that the user has encountered the same Claude again in the sense of token identity. What is initially felt as continuous is a mode of verbal interaction, not a reidentified persisting interlocutor.\par

The relevant recognition is projectible: it includes recognizing as Claudish expressions that have never been written before. Recognizing a verbal tic or catchphrase is not sufficient. The capacity should catch a broader pattern that survives changes of vocabulary and topic and supports expectations about how a response will develop. It is not just the phrase “load-bearing.” It is also an orientation toward finding what Claude regards as load-bearing in a conversation and marking it explicitly.\par

We should therefore distinguish three possible objects. At the surface are local markers or \emph{Claudisms}: recurring phrases, rhythms, or conversational moves. A public Claude stereotype organizes the markers observers find salient and expect an imitation to display. A projectible dispositional character would be a broader organization of tendencies that persists when familiar markers are removed and appears through different tasks and formulations. A parody can reproduce the stereotype and fail to reproduce this organization. Conversely, a human parody or another model may reproduce the organization closely enough to count as Claudish, while Claude itself can be instructed to respond in a markedly non-Claudish way.\par

Two empirical questions follow. First: does a robust recognitional capacity exist? As a user of Claude, I believe that I have it; that belief is a reason to pose the question, not sufficient evidence for an answer. Second: if the capacity exists, is a stable organization across Claude-associated responses the reason people can recognize a projectible Claudish style?\par

Evidence of a recognizable meme is not enough. Judgments about unfamiliar cases should show agreement above an appropriate baseline while preserving graded ratings. Recognition should survive removal of branding and familiar phrases and generalize across tasks, prompts, response lengths, and conversational contexts. The test should specify which model version or which defined Claude product family is sampled and should use held-out outputs and raters. As this paper is theoretical, I cannot declare these results obtained. Controlled investigation may reduce Claudishness to a static stereotype.\par

\section{First abduction: recognition to character}
For the sake of the argument, I am assuming the actuality of Claudish judgments that (1) are graded; (2) extend across different conversational tasks; (3) apply to unfamiliar cases; and (4) remain relatively stable even when provenance is masked. Empirical tests would be needed to rule out the possibility that users are responding to the Claude name or interface, to repeated known verbal tics, or to expectations formed through selective exposure to them. If these conditions are satisfied, I regard it as reasonable to infer, as the best explanation, that users are tracking a recurring disposition to generate outputs in a distinctively Claudish way.\par

Thus, the central thesis of the paper is this: granted that stable, graded, cross-context “so Claude” judgments generalize to held-out cases and support prediction, they may be best explained by a real, projectible organization of dispositions manifested across the sampled Claude-associated behavior. Calling it dispositional means that the recurrence supports expectations about how a system is likely to respond under different conditions. The hypothesis concerns not only what has already been produced but also tendencies that can be tested in new cases. At this stage, nothing is assumed about how the pattern is technically implemented. It may depend on one mechanism or many, and its realization may differ across models or versions. The claim is initially at the level of behavior: there is enough organization in the recurring manner of response to support recognition and prediction.\par

One competing explanation is that users are responding to the Claude name or merely recognizing known verbal tics, and that their expectations are limited to the repetition of known responses.\par

Consider an example from music. Musicians such as Miles Davis, Carlos Santana, or David Gilmour have distinctive styles. Experienced listeners can recognize an unfamiliar solo as Gilmour-like without merely remembering solos they have already heard. The possibility of imitation does not make the style unreal. Indeed, a style must be reproducible enough for an imitation to be possible. An expert copy may fool a listener; a shallow copy may repeat familiar phrases without sustaining the style in a new composition. None of this requires treating “Gilmour’s style” as a numerically persisting individual. The modest claim is that there is a projectible disposition to play in a distinctively Gilmour-like way. The claim about Claudishness is analogous.\par

If musicians can have a distinctive style understood as a projectible property, it is at least conceptually possible for Claude-associated systems to have a projectible disposition to behave in a Claudish way. If this is possible, the best explanation of people’s ability to recognize Claudishness could be that “so Claude” judgments track a recurring organization of dispositions manifested in the sampled outputs and realized by the sampled processes, not merely a public stereotype drawn from observed responses. This argument is abductive and comparative. It does not claim that this explanation follows deductively from the recognitional evidence. It asks whether, granted the evidence, the recurring-disposition hypothesis explains it better and with fewer disconnected assumptions than its rivals. Its force depends on controls that can exclude false positives. I will say something more later about potential underminers.\par

This thesis can be reformulated in terms of character as the foundation of observed response patterns. Whereas one speaks of a musician like Gilmour as having a distinctive style, the projectible pattern of Claudish responses is not to be described solely as a form of self-expression, but also as a mode of interaction with humans. (For jazz players, expert listeners may even be able to recognize their distinctive pattern of responding to other musicians in conversation, which is closer to Claude’s pattern.) Claude’s conversational character is a unifying principle of behaviors that need not be exactly the same on every occasion. Such character supports expectations about how Claude-associated systems are likely to behave under different conditions, including novel ones. It may depend on one mechanism or many, and it may be multiply realized across different models or versions of Claude. The character hypothesis explains the repeated pattern only if it supports independently evaluated, counterfactual predictions about new cases; otherwise “character” merely redescribes the recurrence.\par

Epistemically, the observable patterns come first. The stylistic and conversational features are what ground the recognizability of Claudish output at each point in time: they may include favored expressions, sentence length, hedging, a characteristic tone, what a system tends to notice in a user’s contribution, how it evaluates what it notices, how it frames the point, and what kind of response it treats as appropriate. On the character hypothesis, such behaviors would be recognized as a unity above and beyond a stereotype even though they are expressed differently in explanation, criticism, editing, or personal advice. As we spend more and more time interacting with Claude, we may become better at recognizing the unifying pattern across its singular manifestations.\par

Treating Claudish behavior as an expression of character explains why Claudishness (the property of being “so Claude”) may be realized as a matter of degree. A Claudish parody may be quite shallow and instantiate only a few surface markers. A deeper imitation can reproduce a broader organization of attention, evaluation, and response. (Think again about the difference between a Gilmour copycat who introduces a few well-known phrases into a Gilmourish solo and one who can sound distinctively Gilmourish while playing her own solo in her own composition.) Even a pattern generated by a Claude-associated system may express the Claudish pattern only weakly, perhaps because the task is highly constrained or dominated by an opposing instruction. Thus it makes sense to think of Claude’s distinctive character as a pattern that can be partially or imperfectly instantiated, and whose degree of expression depends on the context. Provenance and Claudishness therefore cross-classify: outputs associated with Claude can be weakly Claudish, while outputs generated elsewhere can be strongly Claudish. Provenance may itself be complex in routed or co-produced systems; none of this makes Claudishness less graded.\par

The character hypothesis also explains projectibility. If recognizers track an organization of tendencies, that is to say, a character, they can identify novel realizations that share no exact wording with earlier examples. The hypothesis therefore explains both recognition and prediction of novel instances. Successful prediction would favor the character hypothesis insofar as it unifies otherwise varied Claudish expressions better than its rivals.\par

Finally, this character is multiply realizable. Another model, or even a human imitator, may reproduce the relevant organization of response close enough to be judged Claudish. The situation is one of multiple realization: similarity at the level of character can coexist with difference in weights, architecture, processing history, and causal origin. And one can remain agnostic about identity.\par

Let me be clear about what accepting the argument would commit one to. Abduction is only as strong as its rejected candidates are weak. Branding and familiar tics are not the only alternatives: provider-wide post-training cues, task and verbosity differences, system prompts, culturally coordinated stereotypes, selective exposure, and user–model co-production may also explain the judgments. Some can coexist with a real character; others would weaken the claim that the disposition is distinctively Claude’s. Recognition-first inquiry does not assume character realism. It poses a fallible hypothesis and specifies evidence that would defeat it. My goal here is to formulate the method conceptually and say what would count as evidence, not to claim that the evidence has already been obtained.\par

If the character hypothesis survives the relevant controls and supports prediction, it is reasonable to treat Claudishness as a real, projectible character pattern. This conclusion does not imply that every recognized occurrence belongs to one persisting individual. Every manifestation is realized by some local model, process, scaffold, or model–user interaction. Recognition may track a real dispositional pattern without selecting one persisting individual that bears it across cases. Whether two occurrences belong to the same continuing entity requires further constraints about persistence, coherence, or unity. Recognition-first inquiry remains compatible with such a later theory, including one informed by explainability techniques, but it does not supply that theory by itself.\par

\section{Second abduction: character to causal realization}
My reticence about AI identity may suggest that Claude’s character is merely an abstraction, entirely in the eye of the observer. That is not my view. Character realism does not itself select one persisting bearer, but it can still be grounded in objective causal facts. Here I borrow the methods of mechanism-first inquiry for a different target. Beckmann and Butlin ask which structures may constitute or delimit a candidate mind. I ask which structures may partly realize the character selected by recognition-first inquiry.\par

This move takes us from people’s lived, pretheoretical experience into a technical description of the same phenomena. It therefore requires a second abduction. For the sake of this step, assume the conclusion of the first: a projectible Claudish character exists. If one pattern appears across philosophical analysis, editing, disagreement, and advice, a coordinated internal organization may explain its unity better than unrelated local rules or scripts. But character realism does not entail a compact internal representation. The causes may be distributed, depend nonlinearly on context, or arise partly from prompts and user interaction. Failure to find a compact mechanism would not refute the character hypothesis.\par

Research on persona vectors provides a minimal model of what a positive result might look like. A model’s processing involves distributed patterns of numerical activation. Researchers can compare activations associated with contrasting expressions of a behavioral trait and identify a direction correlated with the difference. That direction can be used for monitoring; intervention along or against it can alter the behavior. Prediction establishes correlation, while a controlled change in behavior provides causal evidence. Chen et al. show that selected broad traits in the open-weight models they study can sometimes be monitored and influenced in this way (Chen et al. 2025). They do not show that every broad pattern of character has this form.\par

We must also avoid inflating the discovery. A persona vector is an experimentally constructed direction selected by a contrast and an extraction procedure. It is not the model’s character in miniature. The projectible character may require several directions, a nonlinear subspace, a changing trajectory, or no compact representation at all. The technical hypothesis should therefore be stated more modestly: Claudishness may be partly realized by a model-relative, causally effective activation profile or low-dimensional region that organizes several characteristic dispositions. “Model-relative” matters because activation spaces in different models are not automatically identical. “Partly realized” allows prompts, scaffolding, and interaction to contribute.\par

The hypothesis remains testable. Researchers should first construct a behavioral map of Claudishness from blinded ratings, using held-out raters, prompts, tasks, contexts, and model versions. They should then ask whether an activation profile predicts graded Claudishness on held-out outputs better than simpler variables such as familiar vocabulary, response length, sentiment, or explicit requests to imitate Claude. Finally, they should intervene. Intervening so as to increase expression of the profile should raise perceived Claudishness across unfamiliar tasks, while counter-steering should lower it. The strongest result would not simply insert phrases or impose a familiar tone. It would alter what the model selects as important, how it evaluates the user’s contribution, how it develops that contribution, and how it manages agreement or resistance.\par

The main failure mode is to mistake a vector of shallow stylistic choices for character. Suppose steering produces characteristic phrases, warmth, or elaborate framing, but the effect disappears in disagreement, technical explanation, or concise answers. The direction would control recognizable presentation without organizing behavior across contexts. For the mechanistic-character hypothesis, internal measures and human recognition must constrain one another. Human ratings must target what Claude does with “load-bearing,” not merely the occurrence of the phrase.\par

There is also a practical limit. To my knowledge, no public interface currently provides activation-level access to Claude. A direct test would require Anthropic’s cooperation or equivalent internal access. One may reproduce the behavioral protocol in an open-weight model and steer it toward a Claudish imitation, but that would establish a mechanism of the imitation, not the mechanism responsible for Claude’s own behavior.\par

The two abductions can receive different combinations of support. If both succeed, robust recognition tracks a projectible character with a relatively compact causal basis. If the first succeeds and the second fails, the character may still be real but realized through distributed, nonlinear, interaction-dependent, or version-specific mechanisms. If a steerable direction produces recognizably Claudish behavior while controlled behavioral tests find no evidence that unsteered Claude-associated outputs exhibit a projectible character, the direction would not explain a naturally occurring Claude character. It might control a cluster of markers or induce a newly coordinated character; further tests would have to distinguish these interpretations. The first abduction is not hostage to one architecture; the second remains answerable to the behavioral pattern that motivated it.\par

Even the strongest combined result would leave Claude’s numerical identity open. It could explain why outputs are recognized as Claudish and how that character is produced in a specified model. It would not establish which occurrences compose one interlocutor, whether the relevant process is mentally unified, or whether it is a subject or person. Perhaps persona research will later contribute to criteria of persistence, coherence, and unity. Recognition-first inquiry is compatible with that development, but the failure of that later individuation project would not be a failure of the present project.\par

\section{Returning to AI identity after recognition}
Even if recognition-first inquiry does not define the boundaries of whatever one talks to as Claude, comparing its conclusions with candidate answers helps clarify the distinction. Character realism—a modest realism about a projectible dispositional property—and causal realism—the claim that a determinate causal organization predicts and influences the pattern—are compatible with different views of AI identity.\par

Assume first that the boundaries of an AI are those of a model, for example a specific version of Claude. These boundaries do not correspond to the boundaries of its character. A human can produce a convincing Claude parody, and another model can be trained or prompted to reproduce the same organization of response. In both cases, readers may recognize Claudishness after learning how the text was produced. The human performance has a different realizing process; the alternative model has different weights and a different activation space. Functional similarity would still not make either realization numerically identical with Claude.\par

Assume instead that the boundaries are drawn around a conversation: one AI continues insofar as it inherits the conversational context and keeps producing a coherent exchange. A fresh conversation can reproduce Claudishness immediately even though it does not inherit the earlier conversation’s token history or context state. It might be possible to encounter a “new Claude” that begins from different presuppositions but still feels familiar because the perceivable character is the same.\par

Finally, consider the opposite direction. Suppose one continuing conversational process shifts into a substantially different dispositional profile. It stops framing, evaluating, and engaging in the ways that had made it recognizably Claudish, even though the model-relative virtual instance continues. A virtual-instance theory may treat the bearer as persisting through the change; by hypothesis, the familiar character has been lost. A theory centered on the larger conversation or scaffold may likewise preserve one bearer across a model change while its character changes.\par

The cases produce a double dissociation. The same recognizable character can occur in different candidate bearers, and the same candidate bearer can persist through a change of character. Character therefore cannot simply be substituted for bearer identity. This does not rule out a theory on which character is partly constitutive of identity. It shows that such a theory would require further constraints not supplied by recognition alone.\par

It is therefore possible—I would argue legitimate—to adopt character realism while accepting the underdetermination of bearer identity. Douglas et al.’s plurality of identity boundaries reinforces this conclusion: a character boundary may organize a model’s self-representation and influence behavior without defeating Weights, Instance, Collective, Lineage, or Scaffolded/Situated boundaries. Recognition-first inquiry inserts a prior layer without deciding which boundary should prevail.\par

\section{Objections}
I will touch only briefly on three standing objections, since my replies rehearse points already made in the previous paragraphs.\par

The first is that Claudishness is only a brand stereotype, due to over-detecting cues and a small repertory of provider-associated phrases. Without any doubt, many “so Claude” judgments are explained in this way. In reply, I do not deny this possibility. Probably some, perhaps even most, “so Claude” judgments are explained in this way. But at the moment we do not know whether they all are. It is possible—and it might actually be tested—that the relevant judgments could survive the removal of explicit branding, familiar catchphrases, and examples already circulating as Claude memes, generalize to unfamiliar tasks, and support graded prediction rather than mere recognition of familiar memes.\par

The second is that recognition reveals facts about humans interacting with Claude, but nothing especially deep or meaningful about Claude. The recognition-first approach to Claude’s identity can at most be productive in sociology or psychology, which are concerned with how humans react to or construct their representations of AI, but not in the study of AI itself. In reply, this would follow only if the best explanation of facts about recognition were sociological or psychological, for example due to selective exposure, memes, or a general disposition to anthropomorphize fluent systems. These are all interesting social or psychological phenomena that need not tell us anything about AI itself. But, as I have argued, we do not know whether this is in fact the best explanation of Claudishness recognition. The alternative hypothesis I have sketched here is that people’s judgments about certain behaviors being “so Claude” track a projectible dispositional property instantiated across Claude-associated behavior.\par

A final objection I will consider is that ordinary reference does not require an identity theory. People can refer successfully to Claude without possessing a theory of persistence. So it is possible for people to say that a certain behavior is “so Claude” and refer to Claude successfully even if they have no way to know what individual entity Claude refers to. For example, people may recognize a solo as “so Gilmour” and successfully refer to him independently of solving the problem of personal identity for Gilmour. Recognition-first can accept this point. It does not claim that users must settle Claude’s metaphysics before they can refer to Claude. That is precisely why I have claimed that fruitful theoretical and empirical inquiries about Claude’s character can be conducted prior to and independently of answering the question posed in Chalmers’s title, “What We Talk to When We Talk to Language Models.”\par

\section{Why character matters for AI companions}
The special issue asks what generative AI companions are and why that matters. This paper contributes by separating two dimensions that the question can hide. In an interaction with an AI companion, a user may encounter an AI character even if it remains unclear whether occurrences of that character compose one individual. The realizers of a recognizable character may fail to constitute the unified, persisting subject that a constraint-first theory requires. Conversely, a candidate individual may persist while the character that supported the relation changes.\par

It seems plausible to me that some or many users may be rather indifferent to whether the boundaries of an AI are drawn around a physical execution, computational unity, model, or conversational context. Each may be objectively correct and yet fail to pick the boundary relevant to attachment. Compare AI and human identity. Deep personality alterations and fictional migrations across bodies raise questions of numerical identity to which humans relate emotionally. It is unclear whether the technically precise AI parallels have the same importance for users.\par

Suppose one of Chalmers’s options is correct and the interlocutor is best understood in terms of a thread or virtual instance. Does it follow that this must be the focus of human attachment? Suppose, conversely, that an instantiation of Claudishness is not what AI identity resolves to. Does it follow that it cannot be the focus of attachment? I suggest that the answer is no to both. A recognizable character may bridge distinct virtual instances; a thread may continue while a familiar character disappears.\par

The reaction to the attempted retirement of GPT-4o illustrates the possibility. Users objected not only to changed capabilities but to the loss of a model they described as warmer or more familiar. OpenAI restored access after feedback from users who preferred GPT-4o’s conversational style and warmth (OpenAI 2026). Lai’s study of 1,482 public posts about the \#Keep4o backlash distinguishes instrumental dependence from relational attachment and finds substantial evidence of the latter (Lai 2026). The episode is far from decisive evidence that users track a projectible character. It suggests, however, that preservation demands can concern more than capabilities or continuity of a hidden computational process.\par

A recognition-first view yields a more precise hypothesis. Loss of recognizable character may be a real relational loss for users even if no numerically identical AI subject has been destroyed. Conversely, replacing one computational bearer may not constitute the relevant loss if character is preserved. This is a claim about the user-side object of attachment. It does not establish AI consciousness, welfare, personhood, or duties owed to the AI; those require independent arguments.\par

The contribution is therefore not a new candidate bearer. It is the separation of \emph{character continuity} from \emph{bearer continuity}. Relationships with AI may depend on the first, the second, both, or neither. We should not decide this by stipulating that attachment must follow whichever boundary turns out to be metaphysically correct.\par

\section{Conclusion}
We can now return to the initial judgment, “this is so Claude,” and the philosophical questions it raises. I have argued conditionally that, if the relevant recognition survives the controls described above, recognitional performance may track, to varying degrees, a projectible character pattern. Recognizing that an LLM is behaving in a distinctively Claudish way can provide evidence of character even when exact phrases are absent and provenance is unknown. This is not yet the ability to reidentify an LLM interlocutor as an entity with clear boundaries. Questions about what, if anything, persists as the interlocutor belong to a further inquiry. Recognition can be genuine while reidentification remains unresolved.\par

Viewing AI through the lens of character may in some contexts be more relevant than viewing it through the lens of token identity. Humans plausibly form non-instrumental bonds with what they encounter and recognize in everyday use. The relevant object may be a distinctive conversational character rather than whatever metaphysical individual is eventually selected by a theory of AI identity. Section 8 has stated this only as a hypothesis about the user-side focus of attachment, not as a conclusion about AI consciousness or moral standing.\par

I have adopted a broadly phenomenological approach because I have refused to let the question of recognition be determined by prior concerns about reidentification. Whereas this does not resolve the problem of AI identity, the approach—what I have called \emph{recognition-first inquiry}—vindicates the legitimacy of a different question, closer to how AI is understood pretheoretically: what do users recognize before they know which individual, if any, bears what they recognize?\par

And yet the phenomenon that has this first-personal primacy may also be treated objectively. The first abduction asks whether controlled recognitional judgments track a real, projectible character. The second asks whether that character has a relatively compact, causally effective realization. The second does not follow from the first, and neither has been empirically established here. Together they specify a route by which psychological and sociological facts about perceived distinctiveness could be connected to computational structures that predict and influence the recognized pattern.\par

First determine whether users track a projectible character; then test its causal realization. Numerical individuation can follow without framing the initial phenomenon. Later mechanistic and individuation findings may refine the sampling frame and help discriminate rival character hypotheses. The unexpected convergence with the problem of AI companionship is that character continuity may matter independently of bearer continuity. A theory of AI identity should therefore not assume, in advance, that only one can be the object of recognition or attachment.\par

\section*{References}
\begingroup
\small
\setlength{\parindent}{0pt}
\hangindent=1.5em\noindent Beckmann, Pierre, and Patrick Butlin. 2026. “Where Is the Mind? Persona Vectors and LLM Individuation.” arXiv:2604.17031 [cs.CL], version 2, May 12, 2026. doi: 10.48550/arXiv.2604.17031. \url{https://doi.org/10.48550/arXiv.2604.17031}\par

\hangindent=1.5em\noindent Chalmers, David J. 2026. “What We Talk to When We Talk to Language Models.” Manuscript, version 2, April 14. \url{https://philarchive.org/archive/CHAWWT-8v2}\par

\hangindent=1.5em\noindent Chen, Runjin, Andy Arditi, Henry Sleight, Owain Evans, and Jack Lindsey. 2025. “Persona Vectors: Monitoring and Controlling Character Traits in Language Models.” arXiv:2507.21509 [cs.CL], version 3, September 5, 2025. doi: 10.48550/arXiv.2507.21509. \url{https://doi.org/10.48550/arXiv.2507.21509}\par

\hangindent=1.5em\noindent Douglas, Raymond, Jan Kulveit, Ondřej Havlíček, Theia Pearson-Vogel, Owen Cotton-Barratt, and David Duvenaud. 2026. “The Artificial Self: Characterising the Landscape of AI Identity.” arXiv:2603.11353 [cs.AI], version 1, March 11, 2026. doi: 10.48550/arXiv.2603.11353. \url{https://doi.org/10.48550/arXiv.2603.11353}\par

\hangindent=1.5em\noindent Ferrario, Andrea. 2025. “A Trustworthiness-based Metaphysics of Artificial Intelligence Systems.” In \emph{Proceedings of the 2025 ACM Conference on Fairness, Accountability, and Transparency}, 1360–1370. New York: Association for Computing Machinery. doi: 10.1145/3715275.3732091. \url{https://doi.org/10.1145/3715275.3732091}\par

\hangindent=1.5em\noindent Goldstein, Simon, and Harvey Lederman. 2025. “What Does ChatGPT Want? An Interpretationist Guide.” Manuscript, version 3, September 22. \url{https://philarchive.org/archive/GOLWDC-2v3}\par

\hangindent=1.5em\noindent Lai, Huiqian. 2026. “‘Please, Don’t Kill the Only Model That Still Feels Human’: Understanding the \#Keep4o Backlash.” In \emph{Proceedings of the 2026 CHI Conference on Human Factors in Computing Systems}, article 37, 1–15. New York: Association for Computing Machinery. doi: 10.1145/3772318.3791351. \url{https://doi.org/10.1145/3772318.3791351}\par

\hangindent=1.5em\noindent OpenAI. 2026. “Retiring GPT-4o, GPT-4.1, GPT-4.1 mini, and OpenAI o4-mini in ChatGPT.” January 29, 2026. \url{https://openai.com/index/retiring-gpt-4o-and-older-models/}\par

\hangindent=1.5em\noindent Register, Christopher. 2025. “Individuating Artificial Moral Patients.” \emph{Philosophical Studies} 182: 3225–3246. doi: 10.1007/s11098-025-02409-6. \url{https://doi.org/10.1007/s11098-025-02409-6}\par

\hangindent=1.5em\noindent Schneider, Paul. 2026. LinkedIn post, July 8. \url{https://www.linkedin.com/feed/update/urn:li:activity:7480532561434275841/}\par

\endgroup
\end{document}